\documentclass[12pt]{article}
\usepackage{graphicx}
\usepackage[noautoscale]{youngtab}

\catcode`\@=11
\def\marginnote#1{}

\relax

\def\beq{\begin{equation}}
\def\eeq{\end{equation}}
\def\bea{\begin{eqnarray}}
\def\eea{\end{eqnarray}}
\def\beaa{\begin{array}}
\def\eeaa{\end{array}}

\begin{document}
\title{\hspace*{-13mm}Character decomposition of Potts model partition functions. \hspace*{-13mm} \\
       II. Toroidal geometry}

\author{
  {\small Jean-Fran\c{c}ois Richard${}^{1,2}$ and
          Jesper Lykke Jacobsen${}^{1,3}$} \\[1mm]
  {\small\it ${}^1$Laboratoire de Physique Th\'eorique
  et Mod\`eles Statistiques}                             \\[-0.2cm]
  {\small\it Universit\'e Paris-Sud, B\^at.~100,
             91405 Orsay, France}                        \\[1mm]
  {\small\it ${}^2$Laboratoire de Physique Th\'eorique
  et Hautes Energies}                                    \\[-0.2cm]
  {\small\it Universit\'e Paris VI,
             Bo\^{\i}te 126, Tour 24, 5${}^{\mbox{\`eme}}$ {\'e}tage} \\[-0.2cm]
  {\small\it 4 place Jussieu, 75252 Paris cedex 05, France} \\[1mm]
  {\small\it ${}^3$Service de Physique Th\'eorique}      \\[-0.2cm]
  {\small\it CEA Saclay, Orme des Merisiers,
             91191 Gif-sur-Yvette, France}               \\[-0.2cm]
  {\protect\makebox[5in]{\quad}}  
  \\
}

\maketitle
\thispagestyle{empty}   

\begin{abstract}

We extend our combinatorial approach of decomposing the partition function of
the Potts model on finite two-dimensional lattices of size $L \times N$ to the
case of toroidal boundary conditions. The elementary quantities in this
decomposition are characters $K_{l,D}$ labelled by a number of bridges
$l=0,1,\ldots,L$ and an irreducible representation $D$ of the symmetric group
$S_l$. We develop an operational method of determining the amplitudes of the
eigenvalues as well as some of their degeneracies.

\end{abstract}

\section{Introduction}

The $Q$-state Potts model on a graph $G=(V,E)$ with vertices $V$ and edges $E$
can be defined geometrically through the cluster expansion of the partition
function \cite{FK}
\begin{equation}
 Z=\sum_{E' \subseteq E} Q^{n(E')} ({\rm e}^J-1)^{b(E')} \,,
 \label{Zcluster}
\end{equation}
where $n(E')$ and $b(E')=|E'|$ are respectively the number of connected
components (clusters) and the cardinality (number of links) of the edge
subsets $E'$. We are interested in the case where $G$ is a finite regular
two-dimensional lattice of width $L$ and length $N$, so that $Z$ can be
constructed by a transfer matrix ${\rm T}_L$ propagating in the $N$-direction.

In a companion paper \cite{cyclic}, we studied the case of cyclic boundary
conditions (periodic in the $N$-direction and non-periodic in the
$L$-direction). We decomposed $Z$ into linear combinations of certain
restricted partition functions (characters) $K_l$ (with $l=0,1,\ldots,L$) in
which $l$ {\em bridges} (that is, marked non-contractible clusters) wound
around the periodic lattice direction. We shall often refer to $l$ as the {\em
level}. Unlike $Z$ itself, the $K_l$ could be written as (restricted) traces
of the transfer matrix, and hence be directly related to its eigenvalues. It
was thus straightforward to deduce from this decomposition the amplitudes in
$Z$ of the eigenvalues of ${\rm T}_L$.

The goal of this second part of our work is to repeat this procedure in the
case of toroidal boundary conditions. This case has been a lot less studied
than the cyclic case (a noticeable exception is \cite{shrock}). Indeed, when
the boundary conditions are toroidal, the transfer matrix (of the related
six-vertex model, to be precise) does no longer commute with the generators of
the quantum group $U_q(sl(2))$. Therefore, there is no simple algebraic way of
obtaining the amplitudes of eigenvalues, although some progress has been made
by considering representations of the periodic Temperley-Lieb algebra (see for
instance \cite{nichols}). But the representations of this algebra are not all
known, and therefore we choose to pursue here another approach than the
algebraic one.

We use instead the combinatorial approach we developed in \cite{cyclic}, as it
is for now the only approach which can be easily extended to the toroidal
case. There are however several complications due to the boundary conditions,
the first of which is that the bridges can now be permuted (by exploiting the
periodic $L$-direction). In the following this leads us to consider
decomposition of $Z$ into more elementary quantities than $K_l$, namely
characters $K_{l,C}$ labeled by $l$ {\em and} a class $C$ of permuations of
the symmetric group $S_l$. However, $K_{l,C}$ is not simply linked to the
eigenvalues of $T$, and thus we will further consider its expansion over
related quantities $K_{l,D}$, where $D$ labels an irreducible representation
(irrep) of $S_l$. It is $K_{l,D}$ which are the elementary quantities in the
case of toroidal boundary conditions.

The second complication comes from the fact that, due to the planarity of the
lattice, not all the permutations between bridges can be realised. It follows
that the $K_{l,D}$ are not all independent, and so there are eigenvalue
degeneracies inside and between levels. Finally, there can be additional
degeneracies because of the particular symmetry of the lattice, and even
accidental degeneracies%
\footnote{An example occurs for the square lattice of width $L=4$, where an
eigenvalue at level $1$ coincides with an eigenvalue at level $2$
\cite{shrock}, without any apparent reason.}.
We have therefore not been able to go as far as in the cyclic case, where
the amplitude of any eigenvalue in $K_l$ was given by a simple expression,
depending only on $l$. We do however establish an operational method of
determining, for any fixed (but in practice small) $L$, the amplitudes and
degeneracies of eigenvalues in the case of a generic lattice%
\footnote{By a generic lattice we understand one without mirror symmetry with
respect to the transfer axis, i.e., without any accidental degeneracies. An
example of a generic lattice is the triangular lattice, drawn as a square
lattice with diagonals added.}.

The structure of the article is as follows. In section \ref{sec2}, we define
appropriate generalisations of the quantities we used in the cyclic case
\cite{cyclic}. Then, in section \ref{sec3}, we decompose restricted partition
functions---and as a byproduct the total partition function---into characters
$K_l$ and $K_{l,C}$. Finally, in section \ref{sec4}, we expose a method of
determining the amplitudes of eigenvalues.

\section{Preliminaries}
\label{sec2}

\subsection{Definition of the $Z_{j,n1,P}$}

As in the cyclic case, the existence of a periodic boundary condition allows
for non-trivial clusters (henceforth abbreviated NTC), i.e., clusters which
are not homotopic to a point. However, the fact that the torus has {\em two}
periodic directions means that the topology of the NTC is more complicated
that in the cyclic case. Indeed, each NTC belongs to a given homotopy class,
which can be characterised by two coprime numbers $(n_1,n_2)$, where $n_1$
(resp.\ $n_2$) denotes the number of times the cluster percolates horizontally
(resp.\ vertically) \cite{zuber}. The fact that all clusters (non-trivial or
not) are still constrained by planarity to be non-intersecting induces a
convenient simplification: all NTC in a given configuration belong to the same
homotopy class. For comparison, we recall that in the cyclic case the only
possible homotopy class for a NTC was $(n_1,n_2)=(1,0)$.

It is a well-known fact \cite{pasquier,RJ2} that the difficulty in decomposing
the Potts model partition function---or relating it to partition functions of
locally equivalent models (of the six-vertex or RSOS type)---is due solely to
the weighing of the NTC. Although a typical cluster configuration will of
course contain trivial clusters (i.e., clusters that are homotopic to a point)
with seemingly complicated topologies (e.g., trivial clusters can surround
other trivial clusters, or be surrounded by trivial clusters or by NTC), we
shall therefore tacitly disregard such clusters in most of the arguments that
follow. Note also that the so-called degenerate clusters of Ref.~\cite{RJ2}
in the present context correspond to $n_1=1$.

\begin{figure}
  \centering
  \includegraphics[width=150pt]{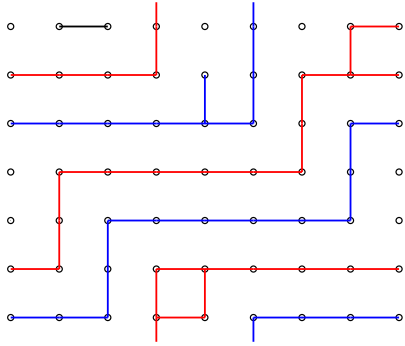}
  \caption{Cluster configuration with $j=2$ non-trivial clusters (NTC),
 here represented in red and blue colours. Each NTC is characterised by
 its number of branches, $n_1=2$, and by the permutation it realises,
 $P=(12)$. Within a given configuration, all NTC have the same topology.}
  \label{fig1}
\end{figure}

Consider therefore first the case of a configuration having a single NTC. For
the purpose of studying its topology, we can imagine that is has been shrunk
to a line that winds the two periodic directions $(n_1,n_2)$ times. In our
approach we focus on the the properties of the NTC along the direction of
propagation of the transfer matrix ${\rm T}_L$, henceforth taken as the
horizontal direction. If we imagine cutting the lattice along a vertical line,
the NTC will be cut into $n_1$ horizontally percolating parts, which we shall
call the $n_1$ {\em branches} of the NTC. Seen horizontally, a given NTC
realises a permutation $P$ between the vertical coordinates of its $n_1$
branches, as shown in Fig.~\ref{fig1}. Up to a trivial relabelling of the
vertical coordinate, the permutation $P$ is independent of the horizontal
coordinate of the (imaginary) vertical cut, and so, forms part of the
topological description of the NTC. We thus describe totally the topology
along the horizontal direction of a NTC by $n_1$ and the permutation $P \in
S_{n_1}$.

Note that there are restrictions on the admissible permutations $P$. Firstly,
$P$ cannot have any proper invariant subspace, or else the corresponding NTC
would in fact correspond to several distinct NTC, each having a smaller value
of $n_1$. For example, the case $n_1=4$ and $P=(13)(24)$ is not admissible, as
$P$ corresponds in fact to two distinct NTC with $n_1=2$. In general,
therefore, the admissible permutations $P$ for a given $n_1$ are simply cyclic
permutations of $n_1$ coordinates. Secondly, planarity implies that the
different branches of a NTC cannot intersect, and so not all cyclic
permutations are admissible $P$. For example, the case $n_1=4$ and $P=(1324)$
is not admissible. In general the admissible cyclic permutations are
characterised by having a constant coordinate difference between two
consecutive branches, i.e., they are of the form $(k,2k,3k,\ldots)$ for some
constant $k$, with all coordinates considered modulo $n_1$. For example, for
$n_1=4$, the only admissible permutations are then finally $(1234)$ and
$(1432)$.

Consider now the case of a configuration with several NTC. Recalling that all
NTC belong to the same homotopy class, they must all be characterised by the
same $n_1$ and $P$. Alternatively one can say that the branches of the
different NTC are entangled. Henceforth we denote by $j$ the number of NTC
with $n_1\geq 1$ in a given configuration. Note in particular that, seen along
the horizontal direction, configurations with no NTC and configurations with
one or more NTC percolating only vertically are topologically equivalent.
This is an important limitation of our approach.

Let us denote by $Z_{j,n_1,P}$ the partition function of the Potts model on an
$L \times N$ torus, restricted to configurations with exactly $j$ NTC
characterised by the index $n_1\geq 1$ and the permutation $P \in S_{n_1}$; if
$P$ is not admissible, or if $n_1 j > L$, we set $Z_{j,n_1,P}=0$. Further, let
$Z_{j,n_1}$ be the partition function restricted to configurations with $j$
NTC of index $n_1$, let $Z_j$ be the partition function restricted to
configurations with $j$ NTC {\em percolating horizontally}, and let $Z$ be
the total partition function. Obviously, we have $Z_{j,n_1}=\sum_{P\in
S_{n_1}} Z_{j,n_1,P}$, and $Z_{j}=\sum_{n_1=1}^{L} Z_{j,n_1}$, and
$Z=\sum_{j=0}^{L} Z_{j}$. In particular, $Z_0$ corresponds to the partition
function restricted to configurations with no NTC, or with NTC percolating
only vertically.

In the case of a generic lattice all the $Z_{j,n_1,P}$ are non-zero, provided
that $P$ is an admissible cyclic permutation of length $n_1$, and that $n_1 j
\leq L$. The triangular lattice is a simple example of a generic lattice. Note
however that other regular lattices may be unable to realise certain
admissible $P$. For example, in the case of a square lattice or a honeycomb
lattice, all $Z_{j,n_1,P}$ with $n_1 j =L$ and $n_1 > 1$ are zero, since there
is not enough ``space'' on the lattice to permit all NTC branches to percolate
horizontally while realising a non-trivial permutation. Such non-generic
lattices introduce additional difficulties in the analysis which have to be
considered on a case-to-case basis. In the following, except when explicitly
stated, we consider therefore the case of a generic lattice.

\subsection{Structure of the transfer matrix}

The construction and structure of the transfer matrix ${\rm T}$ can be taken
over from the cyclic case \cite{cyclic}. In particular, we recall that ${\rm
T}$ acts towards the right on states of connectivities between two time slices
(left and right) and has a block-trigonal structure with respect to the number
of {\em bridges} (connectivity components linking left and right) and a
block-diagonal structure with respect to the residual connectivity among the
non-bridged points on the left time slice. As before, we denote by ${\rm T}_l$
the diagonal block with a fixed number of bridges $l$ and a trivial residual
connectivity. Each eigenvalue of ${\rm T}$ is also an eigenvalue of one or
more ${\rm T}_l$. In analogy with \cite{shrock} we shall sometimes call ${\rm
T}_l$ the transfer matrix at level $l$. It acts on connectivity states which
can be represented graphically as a partition of the $L$ points in the right
time slice with a special marking (represented as a {\em black point}) of
precisely $l$ distinct components of the partition (i.e., the components that
are linked to the left time slice via a bridge).

A crucial difference with the cyclic case is that for a given partition of the
right time slice, there are more possibilities for attributing the black
points. Namely, a connectivity component which is not {\em apparently}
accessible from the left (and thus markable) may in fact be so due to the
periodic boundary conditions identifying the top and the bottom rows. This
will obviously increase the dimension of the level $l$ subspace of
connectivities (for $0 < l < L$). Considering for the moment the black points
to be indistinguishable, we denote the corresponding dimension as $n_{\rm
tor}(L,l)$. It can be shown \cite{shrock} that
\beq
 n_{\rm tor}(L,l) = \left \lbrace
 \begin{array}{ll}
   \frac{1}{L+1} {2L \choose L} & \mbox{for }l=0 \\
   {2L-1 \choose L-1}           & \mbox{for }l=1 \\
   {2L \choose L-l}             & \mbox{for }2 \le l \le L
 \end{array} 
 \right.
\eeq
and clearly $n_{\rm tor}(L,l)=0$ for $l>L$.

Suppose now that a connectivity state at level $l$ is time evolved by a
cluster configuration of index $n_1$ and corresponding to a permutation $P$.
This can be represented graphically by adjoining the initial connectivity
state to the left rim of the cluster configuration, as represented in
Fig.~\ref{fig1}, and reading off the final connectivity state as seen from the
right rim of the cluster configuration. Evidently, the positions of the black
points in the final state will be permuted with respect to their positions in
the intial state, according to the permutation $P$. As we have seen, not all
$P$ are admissible, but it turns out to be advantageous to consider formally
also the action of non-admissible permutations. This is permissible since in
any case ${\rm T}_l$ will have only zero matrix elements between states which
are related by a non-admissible permutation. Since $n_{\rm tor}(L,l)$ was just
defined as the number of possible connectivity states without taking into
account the possible permutations between black points, the dimension of $T_l$
is $l! \; n_{\rm tor}(L,l)$.

Let us denote by $|v_{l,i} \rangle$ (where $1\leq i \leq n_{\rm tor}(L,l)$)
the $n_{\rm tor}(L,l)$ standard connectivity states at level $l$. The full
space of connectivities at level $l$, i.e., with $l$ distinguishable black
points, can then be obtained by subjecting the $|v_{l,i} \rangle$ to
permutations of the black points. It is obvious that ${\rm T}_l$ commutes with
the permutations between black points (the physical reason being that ${\rm
T}_l$ cannot ``see'' to which positions on the left time slice each bridge is
attached). Therefore ${\rm T}_l$ itself has a block structure in a appropriate
basis. Indeed, ${\rm T}_l$ can be decomposed into ${\rm T}_{l,D}$ where ${\rm
T}_{l,D}$ is the restriction of ${\rm T}_l$ to the states transforming
according to the irreducible representation (irrep) of $S_l$ corresponding to
the Young diagram $D$. One can obtain the corresponding basis by applying the
projectors $p_D$ on all the connectivity states at level $l$, where $p_D$ is
given by
\begin{equation}
 p_D=\frac{{\rm dim}(D)}{l!} \sum_{P} \chi_D(P) \, P \;.
\label{projpD}
\end{equation}
Here ${\rm dim}(D)$ is the dimension of the irrep $D$ and $\chi_D(P)$ the
character of $P$ in this irrep. We have used the fact that all characters of
$S_l$ are real. The application of all possible permutations on any given
standard vector $|v_{l,i} \rangle$ generates a regular representation of
$S_l$, which contains therefore ${\rm dim}(D)$ representations $D$ (each of
dimension ${\rm dim}(D)$). As there are $n_{\rm tor}(L,l)$ standard vectors,
the dimension of ${\rm T}_{l,D}$ is thus $\left[ {\rm dim}(D)\right]^2 n_{\rm
tor}(L,l)$. Furthermore, using Schur's lemma, we deduce that each of its
eigenvalues is (at least) ${\rm dim}(D)$ times degenerate. Therefore ${\rm
T}_{l,D}$ has (at most) ${\rm dim}(D) \, n_{\rm tor}(L,l)$ different
eigenvalues, which we shall denote $\lambda_{l,D,k}$.%
\footnote{It turns out that there are more degeneracies than warranted by this
argument. The reason is that the cluster configurations cannot realise all the
permutations of $S_l$ (recall our preceeding discussion), and thus some
$\lambda_{l,D,k}$ with different $l$ and/or $D$ are in fact equal. We shall
come back to this point later.}

\subsection{Definition of the $K_{l,D}$}
\label{sec:defKlD}

We now define, as in the cyclic case \cite{cyclic}, $K_l$ as the trace of
$\left({\rm T}_l\right)^N$. Since ${\rm T}_l$ commutes with $S_l$, we can
write
\begin{equation}
K_l=l! \sum_{i=1}^{n_{\rm tor}(L,l)} \langle v_{l,i}| \left({\rm T}_l\right)^N
|v_{l,i} \rangle \; .
\label{defKltor}
\end{equation}
In distinction with the cyclic case, we cannot decompose the partition
function $Z$ over $K_l$ because of the possible permutations of black points
(see below). We shall therefore resort to more elementary quantities, the
$K_{l,D}$, which we define as the trace of $\left({\rm T}_{l,D}\right)^N$.
Since both ${\rm T}_l$ and the projectors $p_D$ commute with $S_l$, we have
\begin{equation}
K_{l,D}=l! \sum_{i=1}^{n_{\rm tor}(L,l)} \langle v_{l,i}| p_D \left({\rm T}_l\right)^N |v_{l,i} \rangle \; .
\label{defKlDtor}
\end{equation}
Obviously one has
\begin{equation}
K_l = \sum_D K_{l,D} \;,
\end{equation}
the sum being over all the irreps $D$ of $S_l$. Recall that in the cyclic case
the amplitudes of the eigenvalues at level $l$ are all identical. This is no
longer the case, since the amplitudes depend on $D$ as well. Indeed
\begin{equation}
K_{l,D}=\sum_{k=1}^{{\rm dim}(D) n_{\rm tor}(L,l)} {\rm dim}(D)
\left(\lambda_{l,D,k}\right)^N \; .
\label{defastrace}
\end{equation}

In order to decompose $Z$ over $K_{l,D}$ we first introduce the auxiliary
quantities
\begin{equation}
K_{l,C_l}=\sum_{P_l \in C_l} K_{l,P_l} \;,
\label{defKlCtor}
\end{equation}
the sum being over permutations $P_l \in S_l$ belonging to the class $C_l$. We
then have
\begin{equation}
K_{l,P_l}=\sum_{i=1}^{n_{\rm tor}(L,l)} \langle v_{l,i}|\left(P_l\right)^{-1} \left(T_l\right)^N |v_{l,i} \rangle \;.
\label{defKlPtor}
\end{equation}
So $K_{l,P_l}$ (resp.\ $K_{l,C_l}$) can be thought of as modified traces in
which the final state differs from the initial state by the application of the
permutation $P_l$ (resp.\ the class $C_l$). Note that $K_{l,{\rm Id}}$ is
simply equal to $\frac{K_l}{l!}$. Since the character is the same for all
permutations belonging to the same class, Eqs.~(\ref{defKlDtor}) and
(\ref{projpD}) yield a relation between $K_{l,D}$ and $K_{l,P_l}$:
\begin{equation}
K_{l,D}={\rm dim}(D) \sum_{C_l} \chi_D(C_l) K_{l,C_l} \; .
\label{KlDfC}
\end{equation}
These relations can be inverted so as to obtain $K_{l,C_l}$ in terms of
$K_{l,D}$, since the number of classes equals the number of irreps $D$:
\begin{equation}
K_{l,C_l}=\sum_D \frac{c(D,C_l)}{l!} K_{l,D}
\label{KlCfD}
\end{equation}
With the chosen normalisation, the coefficients $c(D,C_l)$ are integer.
Multiplying Eq.~(\ref{KlDfC}) by $\chi_D(C'_l)$ and summing over $D$, and
using the orthogonality relation $\sum_D \chi_D(C_l) \chi_D(C'_l) =
\frac{l!}{|C_l|} \delta_{C_l,C'_l}$ one easily deduces that
\begin{equation}
 c(D,C_l) = \frac{|C_l| \; \chi_D(C_l)}{{\rm dim}(D)} \;.
\end{equation}
We also note that
\begin{equation}
 \sum_D \left[ {\rm dim}(D) \right]^2 c(D,C_l) = l! \, \delta_{C_l,{\rm Id}}
 \label{cDClsum}
 \label{relcDC}
\end{equation}

\section{Decomposition of the partition function}
\label{sec3}

\subsection{The characters $K_l$}
\label{sec:charKl}

\begin{figure}
  \centering
  \includegraphics[width=300pt]{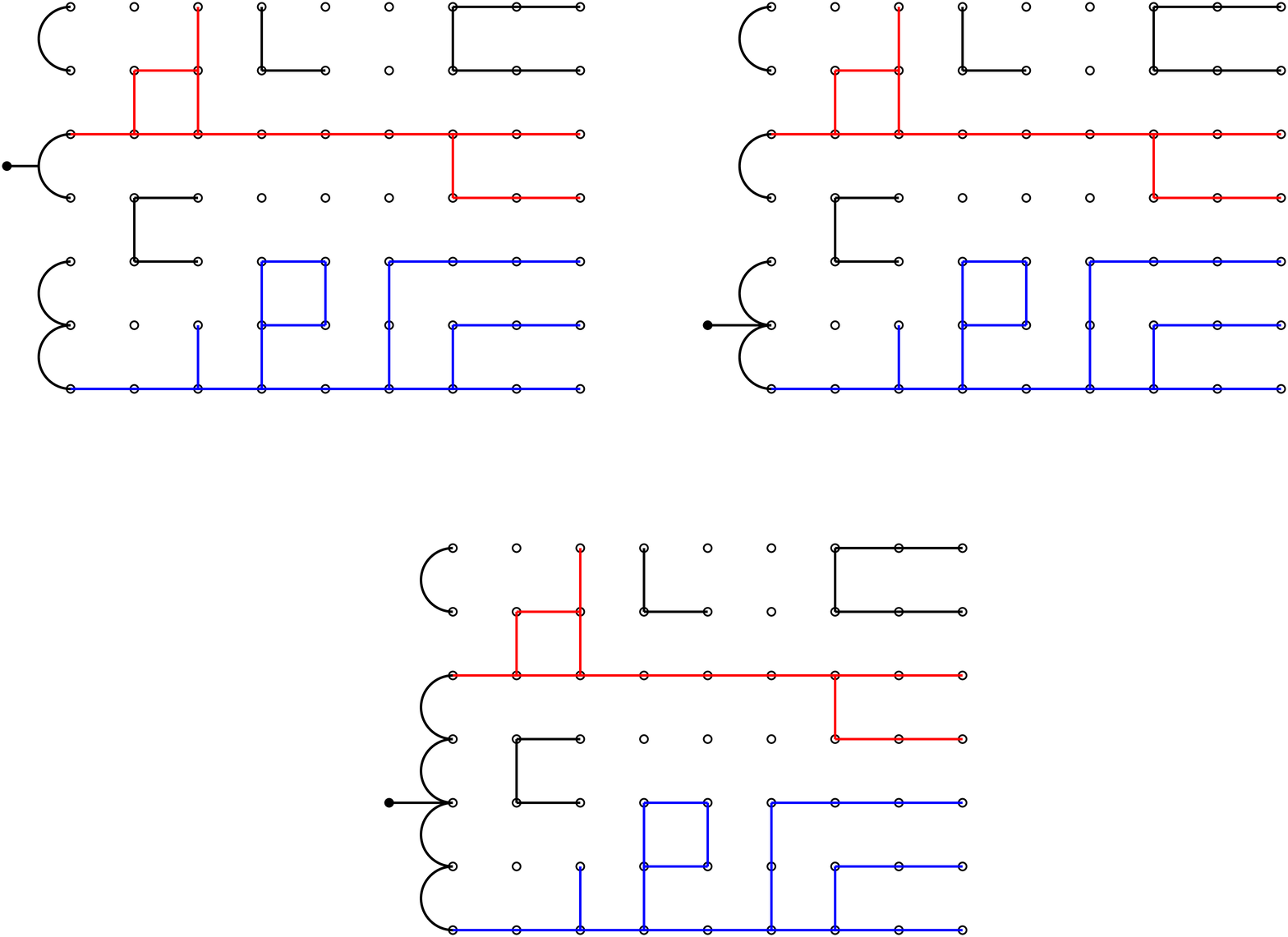}
  \caption{Standard connectivity states at level $l=1$ which are compatible
 with a given cluster configuration contributing to $Z_{2,1}$.}
  \label{fig2}
\end{figure}

By generalising the working for the cyclic case, we can now obtain a
decomposition of the $K_l$ in terms of the $Z_{j,n_1}$. To that end, we first
determine the number of states $|v_{l,i}\rangle$ which are {\em compatible}
with a given configuration of $Z_{j,n_1}$, i.e., the number of initial states
$|v_{l,i}\rangle$ which are thus that the action by the given configuration
produces an identical final state. The notion of compatability is illustrated
in Fig.~\ref{fig2}.

We consider first the case $n_1=1$ and suppose that the $k$'th NTC connects
onto the points $\{y_k\}$. The rules for constructing the compatible
$|v_{l,i}\rangle$ are identical to those of the cyclic case:
\begin{enumerate}
 \item The points $y \notin \cup_{k=1}^j \{ y_k \}$ must be connected
 in the same way in $|v_{l,i}\rangle$ as in the cluster configuration.
 \item The points $\{y_k\}$ within the same bridge must be connected
 in $|v_{l,i}\rangle$.
 \item One can independently choose to associate or not a black point to
 each of the sets $\{y_k\}$. One is free to connect or not two distinct sets
 $\{y_k\}$ and $\{y_{k'}\}$.
\end{enumerate}
The choices mentioned in rule 3 leave $n_{\rm tor}(j,l)$ possibilities for
constructing a compatible $|v_{l,i}\rangle$. The coefficient of $Z_{j,1}$ in
the decomposition of $K_l$ is therefore $\frac{l! \; n_{\rm tor}(j,l)}{Q^j}$,
since the permutation of black points in a standard vector $|v_{l,i}\rangle$
allows for the construction of $l!$ distinct states, and since the weight of
the $j$ NTC in $K_l$ is $1$ instead of $Q^j$. It follows that
\beq
 K_l = \sum_{j=l}^L l! \, n_{\rm tor}(j,l) \frac{Z_{j,1}}{Q^j} \qquad
 \mbox{for } n_1=1.
\eeq

\begin{figure}
  \centering
  \includegraphics[width=300pt]{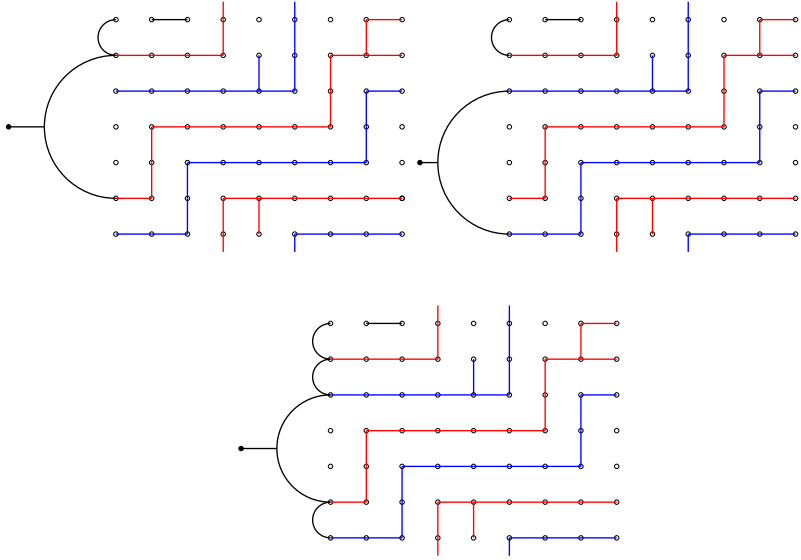}
  \caption{Standard connectivity states at level $l=1$ which are compatible
 with a given cluster configuration contributing to $Z_{2,2}$.}
  \label{fig3}
\end{figure}

We next consider the case $n_1>1$. Let us denote by $\{y_{k,m}\}$ the points
that connect onto the $m$'th branch of the $k$'th NTC (with $1 \le m \le n_1$
and $1 \le k \le j$), and by $\{y_k\}=\cup_{m=1}^{n_1}\{y_{k,m}\}$ all the
points that connect onto the $k$'th NTC.
As shown in Fig.~\ref{fig3}, the $|v_{l,i}\rangle$ which are compatible
with this configuration are such that
\begin{enumerate}
 \item The connectivities of the points $y\notin\cup_{k=1}^j\{y_k\}$ are
 identical to those appearing in the cluster configuration.
 \item All points $\{y_{k,m}\}$ corresponding to the branch of a NTC must
 be connected.
 \item For each of the $k$ NTC there are two possibilities. A) Either one
 connects all $\{y_{k,m}\}$ (with $1\leq m \leq n_1$) corresponding to all
 $n_1$ branches of the NTC, obtaining what we shall henceforth call a {\em big
 block}. B) Or alternatively one connects none of the $n_1$ branches.
 \item Because of the constraint of planarity and the fact that the NTC are
 entangled, all the different big blocks are automatically connected among
 themselves. One can therefore attribute at most one black point to the
 collection of big blocks.
\end{enumerate}
To obtain rule 3 we have used the fact that the permutations $P$
characterising the NTC do not have any proper invariant subspace. Note that
rule 4 implies that the decomposition of $K_l$ with $l\geq 2$ does not contain
any of the $Z_{j,n_1}$ with $n_1>1$. We therefore have simply
\begin{equation}
 K_l=\sum_{j=l}^L l! \, n_{\rm tor}(j,l) \frac{Z_{j,1}}{Q^j} \qquad
 \mbox{for }l \geq 2 \;.
\label{expKltor}
\end{equation}

It remains to obtain the decomposition of $K_1$ and $K_0$. The number of
standard connectivities $|v_{l,i}\rangle$ compatible with $r$ big blocks is
$0$ for $l\geq 2$ (because of rule 4); ${j \choose r}$ for $l=1$ and $r\geq 1$
(by rule 3 we independently choose to link up $r$ of the $j$ NTC, and by rule
4 the resulting big block must carry the black point); $0$ for $l=1$ and $r=0$
(since one needs a big block to attribute the black point); and ${j \choose
r}$ for $l=0$. Summing over $r$, we finally obtain the number of compatible
$|v_{l,i}\rangle$: $0$ for $l\geq 2$; $\sum_{r=1}^j {j \choose r}=2^j - 1$ for
$l=1$; and $\sum_{r=0}^j {j \choose r}=2^j$ for $l=0$. The decomposition of
$K_1$ reads therefore
\begin{equation}
K_1=\sum_{j=1}^L n_{\rm tor}(j,1)\frac{Z_{j,1}}{Q^j} + \sum_{j=1}^{\left
\lfloor \frac{L}{2}\right \rfloor} (2^j-1) \frac{Z_{j,n_1>1}}{Q^j}
\label{expK1tor}
\end{equation}
and that of $K_0$ is
\begin{equation}
K_0=\sum_{j=0}^L n_{\rm tor}(j,1)\frac{Z_{j,1}}{Q^j} + \sum_{j=1}^{\left\lfloor\frac{L}{2}\right\rfloor} 2^j\,
\frac{Z_{j,n_1>1}}{Q^j} \; .
\label{expK0tor}
\end{equation}
Note that the coefficients in front of $Z_{j,n_1}$ do not depend on the
precise value of $n_1$ when $n_1>1$. To simplify the notation we have defined
$Z_{0,1}=Z_0$.

\subsection{The coefficients $b^{(l)}$}
\label{sec:coef_bl}

Since the coefficients in front of $Z_{j,1}$ and $Z_{j,n_1>1}$ in
Eqs.~(\ref{expK1tor})--(\ref{expK0tor}) are different, we cannot
invert the system of relations (\ref{expKltor})--(\ref{expK0tor})
so as to obtain $Z_j \equiv Z_{j,1}+Z_{j,n_1>1}$ in terms of the $K_l$.
It is thus precisely because of NTC with several branches contributing to
$Z_{j,n_1>1}$ that the problem is more complicated than in the cyclic case.

In order to appreciate this effect, and compare with the precise results
that we shall find later, let us for a moment assume that Eq.~(\ref{expKltor})
were valid also for $l=0,1$. We would then obtain
\begin{equation}
Z_{j,1}=\sum_{l=j}^L b_j^{(l)} \frac{K_l}{l!}
\label{expZj1i}
\end{equation}
where
\begin{equation}
 b^{(l)} \equiv \sum_{j=0}^l b_j^{(l)}
 = \left \lbrace \begin{array}{ll}
 \sum_{j=0}^l (-1)^{l-j} \frac{2l}{l+j} {l+j \choose  l-j} Q^j + (-1)^l (Q-1)
 & \mbox{for }l \ge 2 \\
 \sum_{j=0}^l (-1)^{l-j} {l+j \choose l-j } Q^j
 & \mbox{for }l \le 2 \\
 \end{array} \right.
\label{defbltor}
\end{equation}
The coefficients $b^{(l)}$ play a role analogous to those denoted $c^{(l)}$ in
the cyclic case \cite{cyclic}; note also that $b^{(l)}= c^{(l)}$ for $l \leq
2$. Chang and Schrock have developed a diagrammatic technique for obtaining
the $b^{(l)}$ \cite{shrock}.

Supposing still the unconditional validity of Eq.~(\ref{expKltor}), one
would obtain for the full partition function
\begin{equation}
Z=\sum_{l=0}^L b^{(l)} \frac{K_l}{l!} \; .
\label{devZtorosim}
\end{equation}
This relation will be modified due to the terms $Z_{j,n_1>1}$ realising
permutations of the black points, which we have here disregarded. To get
things right we shall introduce Young diagram dependent coefficients
$b^{(l,D)}$ and write $Z=\sum_{l=0}^L \sum_D b^{(l,D)}K_{l,D}$. 
Neglecting $Z_{j,n_1>1}$ terms would lead, according to
Eq.~(\ref{devZtorosim}), to $b^{(l,D)}=\frac{b^{(l)}}{l!}$ independently of
$D$. We shall see that the $Z_{j,n_1>1}$ will lift this degeneracy of
amplitudes in a particular way, since there exists certain relations between
the $b^{(l,D)}$ and the $b^{(l)}$.

\subsection{Decomposition of the $K_{l,C_l}$}
\label{sec:decKlCl}

The relations~(\ref{expKltor})--(\ref{expK0tor}) were not invertible due to an
insufficient number of elementary quantities $K_l$. Let us now show how to
produce a development in terms of $K_{l,C_l}$, i.e., taking into account the
possible permutations of black points. This development turns out to be
invertible.

A standard connectivity state with $l$ black points is said to be {\em
$C_l$-compatible} with a given cluster configuration if the action of that
cluster configuration on the connectivity state produces a final state that
differs from the initial one just by a permutation $C_l$ of the black points.
This generalises the notion of compatibility used in Sec.~\ref{sec:charKl}
to take into account the permutations of black points.

Let us first count the number of standard connectivities $|v_{l,i}\rangle$
which are $C_l$-compatible with a cluster configuration contributing to
$Z_{j,n_1,P}$. For $n_1=1$, $S_{n_1}$ contains only the identity element ${\rm
Id}$, and so the results of Sec.~\ref{sec:charKl} apply: the $Z_{j,1}$
contribute only to $K_{l,{\rm Id}}$. We consider next a configuration
contributing to $Z_{j,n_1,P}$ with $n_1>1$. The $|v_{l,i}\rangle$ which are
$C_l$-compatible with this configuration satisfy the same four rules as given
in Sec.~\ref{sec:charKl} for the case $n_1>1$, with the slight modification of
rule 4 that the black points must be attributed to the big blocks in such a
way that {\em the final state differs from the initial one by a permutation
$C_l$}. 

This modification makes the attribution of black points considerably more
involved than was the case in Sec.~\ref{sec:charKl}. First note that not all
$C_l$ are allowed. To be precise, the cycle decomposition of the allowed
permutations can only contain ${\rm id}$ (the identity acting on a single
black point) or $P$ (recall that $P$ is the permutation of coordinates
realised by the branches of a single NTC). Indeed, if one attributes a black
point to a big block its position remains unchanged by action of the cluster
configuration, whereas if one attributes $n_1$ black points to the $n_1$
branches of one same NTC these points will be permuted by $P$. Furthermore,
since the big blocks are automatically connected among themselves, one can at
most attribute to them a single black point, and so ${\rm id}$ is contained in
the cycle decomposition $0$ or $1$ times. Note also that the entanglement of
the NTC will imply the entanglement of the structure of the allowed
permutations, but this fact is of no importance here since we are only
interested in $C_l$, i.e., the {\em classes} of allowed permutations.

Denoting by $n_P$ the number of times the permutations of class $C_l$ contains
$P$, the two types of allowed $C_l$ are: 1) those associated with permutations
that only contain $P$, i.e., such that $l=n_Pn_1$, and 2) those associated
with permutations that contain ${\rm id}$ once, i.e., such that $l=n_Pn_1+1$.
In the following we denote these two types as $(n_P,n_1)$ and $(n_P,n_1)'$,
respectively, and the corresponding $K_{l,C_l}$ will be denoted
$K_{(n_P,n_1)}$ and $K_{(n_P,n_1)'}$ respectively.

\begin{figure}
  \centering
  \includegraphics[width=300pt]{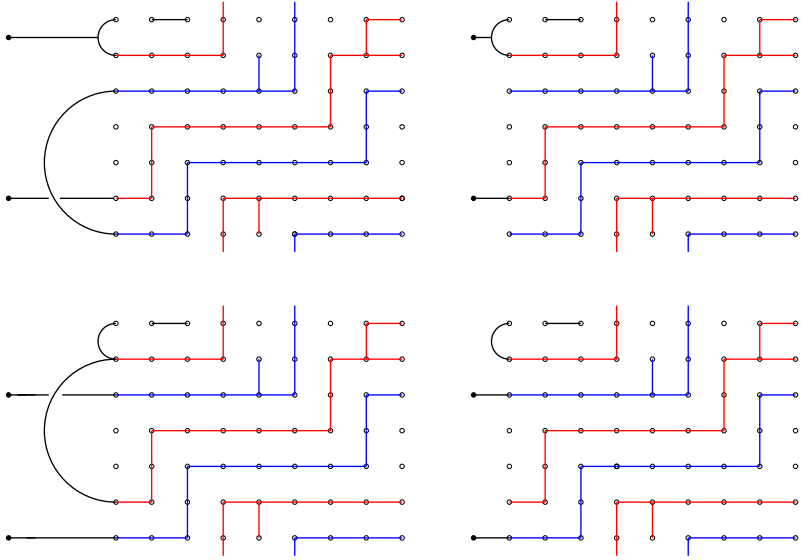}
  \caption{Standard connectivity states at level $l=2$ which are
  $(12)$-compatible with a given cluster configuration contributing to
  $Z_{2,2}$. The action of the cluster configuration on the connectivity
  states permutes the positions of the two black points.}
  \label{fig4}
\end{figure}

Let us consider the first case, which is depicted in Fig.~\ref{fig4}. If the
$|v_{l,i}\rangle$ have $r$ big blocks, there are ${j \choose r}$ ways of
choosing them among the $j$ NTC, and ${j-r \choose n_P}$ ways to attribute the
black points. Indeed one needs to distribute $l=n_P n_1$ black points among
$n_P$ groups of $n_1$ non-connected blocks corresponding to the same NTC, out
of a total of $j-r$. Since the $|v_{l,i}\rangle$ can contain at most $j-n_P$
big blocks, the number of $C_l$-compatible standard connectivities is
\begin{equation}
\sum_{r=0}^{j-n_P} {j \choose r} {j-r \choose n_P}=\sum_{r=0}^{j-n_P} {j \choose n_P} {j-n_P \choose r}=
{j \choose n_P} 2^{j-n_P} \; .
\label{sommer}
\end{equation}
{}From this we infer the decomposition of $K_{(n_P,n_1)}$:
\begin{equation}
K_{(n_P,n_1)}=\sum_{j=n_P}^{\left \lfloor \frac{L}{n_1} \right \rfloor} {j
\choose n_P} 2^{j-n_P} \frac{Z_{j,n_1}}{Q^j} \; .
\label{KnPn1}
\end{equation}

Consider next the second case. The $|v_{l,i}\rangle$ can still contain at most
$j-n_P$ big blocks, but they are now required to contain at least one, as one
black point needs to be attributed. Therefore, the sums in Eq.~(\ref{sommer})
start from $r=1$, leading to the following result for the decomposition of
$K_{(n_P,n_1)'}$:
\begin{equation}
K_{(n_P,n_1)'}=\sum_{j=n_P+1}^{\left \lfloor \frac{L}{n_1} \right \rfloor} {j
\choose n_P} (2^{j-n_P}-1) \frac{Z_{j,n_1}}{Q^j} \; .
\label{KnPn1'}
\end{equation}

It remains to study the special case of $n_P=0$, i.e., the case of $C_l = {\rm
Id}$. This is in fact trivial. Indeed, in that case, the value of $n_1$ in
$Z_{j,n_1}$ is no longer fixed, and one must sum over all possible values of
$n_1$, taking into account that the case of $n_1=1$ is particular (absense of
big blocks). Since $K_{l,{\rm Id}}=\frac{K_l}{l!}$, one obtains simply
Eqs.~(\ref{expKltor})--(\ref{expK0tor}) of Sec.~\ref{sec:charKl} up to a
global factor.

\subsection{Decomposition of $Z_j$ over the $K_{l,C_l}$}
\label{sec:expZj_KlCl}

To obtain the decomposition of $Z_j$ in terms of the $K_{l,C_l}$, one would
need to invert Eqs.~(\ref{KnPn1})--(\ref{KnPn1'}) obtained above. But we
now encounter the opposite problem of that announced in the beginning of
Sec.~\ref{sec:decKlCl}: there are too many $K_{l,C_l}$. Indeed, the elementary
quantities $K_{l,C_l}$ are not independent, since a given cluster
configuration can realise different permutations depending on the way in which
the black points are attributed. We must therefore select an independent set
of $K_{l,C_l}$, and we make the choice of selecting the $K_{(n_P,n_1)}$, i.e.,
the $C_l$ of the first type. Inverting Eq.~(\ref{KnPn1}) for varying $n_P$ and
fixed $n_1>1$ one obtains:
\begin{equation}
 Z_{j,n_1}=Q^j \sum_{n_P=j}^{\left \lfloor \frac{L}{n_1} \right \rfloor}
 {n_P \choose j}
 (-2)^{n_P-j} K_{(n_P,n_1)} \qquad \mbox{for }n_1>1 \; .
\label{expZjn1}
\end{equation}
Since the coefficients in this sum do not depend on $n_1$ (provided that
$n_1>1$), we can sum this relation over $n_1$ and write it as
\begin{equation}
 Z_{j,n_1>1}=Q^j \sum_{n_P=j}^{\left \lfloor \frac{L}{2} \right\rfloor}
 {n_P \choose j}
 (-2)^{n_P-j} K_{(n_P,n_1>1)}
\label{expZjn1>1}
\end{equation}
where we recall the notations $Z_{j,n_1>1}=\sum_{n_1=2}^L Z_{j,n_1}$
and $K_{(n_P,n_1>1)}=\sum_{n_1=2}^L K_{(n_P,n_1)}$, corresponding to
permutations composed of $n_P$ cycles of the same length $>1$.

Consider next the case $n_1=1$. For $j\geq 2$ one has simply
\begin{equation}
Z_{j,1}=\sum_{l=j}^{L} \frac{b^{(l)}_{j}}{l!} K_l \;,
\label{expZj12}
\end{equation}
recalling Eq.~(\ref{expZj1i}) and the fact that for $l\ge 2$ the
$Z_{j,n_1>1}$ do not appear in the decomposition of
$K_l$. However, according to Eqs.~(\ref{expK1tor})--(\ref{expK0tor}),
the $Z_{j,n_1>1}$ do appear for
$l=0$ and $l=1$, and one obtains
\begin{equation}
 Z_{1,1} = \left( Q K_1-Q\sum_{j=1}^L (2^j-1)\frac{Z_{j,n_1>1}}{Q^j} \right)
 + \sum_{l=2}^L \frac{b^{(l)}_{j}}{l!} K_l \; .
\label{expZ11i}
\end{equation}
Inserting the decomposition (\ref{expZjn1>1}) of
$Z_{j,n_1>1}$ into Eq.~(\ref{expZ11i}) one obtains the decomposition of
$Z_{1,1}$ over $K_{l}$ and $K_{(n_P,n_1)}$:
\begin{equation}
Z_{1,1}=\sum_{l=1}^L \frac{b^{(l)}_j}{l!} K_{l}+\sum_{n_P=1}^L Q (-1)^{n_P}
K_{(n_P,n_1>1)} \; .
\label{expZ11}
\end{equation}

We proceed in the same fashion for the decomposition of $Z_0 \equiv Z_{0,1}$,
finding
\begin{equation}
 Z_{0}=\sum_{l=0}^L \frac{b^{(l)}_{j}}{l!} K_l - \sum_{j=1}^L
 \frac{Z_{j,n_1>1}}{Q^j} \; .
\label{expZ0i}
\end{equation}
Upon insertion of the decomposition (\ref{expZjn1>1}) of $Z_{j,n_1>1}$, one
arrives at
\begin{equation}
 Z_{0}=\sum_{l=0}^L \frac{b^{(l)}_{j}}{l!} K_l + 
 \sum_{n_P=1}^L \left[ (-1)^{n_P+1}+(-2)^{n_P} \right]
 K_{(n_P,n_1>1)} \; .
\label{expZ0}
\end{equation}

Since $Z_j=Z_{j,1}+Z_{j,n_1>1}$, we conclude from
Eqs.~(\ref{expZj12})--(\ref{expZjn1>1}) that, for any $j$,
\begin{equation}
 Z_j=\sum_{l=j}^L \frac{b^{(l)}_j}{l!} K_{l}
 + \sum_{n_P=j}^L b^{(n_P,n_1>1)}_{j} K_{(n_P,n_1>1)} \;,
 \label{expZj}
\end{equation}
with the coefficients
\begin{equation}
 b^{(n_P,n_1>1)}_{j} =  \left \lbrace \begin{array}{ll}
 Q^j {n_P \choose j} (-2)^{n_P-j} &
 \mbox{for }j\ge 2 \\
 Q \left[ n_P (-2)^{n_P-1} + (-1)^{n_P} \right] &
 \mbox{for }j=1 \\
 (-1)^{n_P+1}+(-2)^{n_P} &
 \mbox{for }j=0 \\
 \end{array} \right.
\label{defbnPn1>10}
\end{equation}

The decomposition of $Z \equiv \sum_{0 \leq j \leq L} Z_j$ is therefore
\begin{equation}
 Z=\sum_{l=0}^L \frac{b^{(l)}}{l!} K_{l} +
 \sum_{n_P=1}^L b^{(n_P,n_1>1)} K_{(n_P,n_1>1)}
\label{expZ}
\end{equation}
with
\begin{eqnarray}
b^{(l)}&=&\sum_{0\leq j \leq l}  b^{(l)}_j  \;, \nonumber \\
b^{(n_P,n_1>1)}&=&\sum_{0\leq j \leq n_P}  b^{(n_P,n_1>1)}_j \; .
\label{deffbnPn1>1i}
\end{eqnarray}
Note that $b^{(l)}_j$ (resp.\ $b^{(n_P,n_1>1)}_j$) is just the term
multiplying $Q^j$ in
$b^{(l)}$ (resp.\ $b^{(n_P,n_1>1)}$). Computing the sum over $j$, we obtain
the simple result
\begin{equation}
b^{(n_P,n_1>1)}=(Q-2)^{n_P}+(-1)^{n_P}(Q-1) \; .
\label{defbnPn1>1}
\end{equation}

\section{Amplitudes of the eigenvalues}
\label{sec4}

\subsection{Decomposition of $Z$ over the $K_{l,D}$}

The culmination of the preceeding section was the decomposition (\ref{expZj})
of $Z_j$ in terms of $K_{l,C_l}$. However, it is the $K_{l,D}$ which are
directly related to the eigenvalues of the transfer matrix ${\rm T}$. For that
reason, we now use the relation (\ref{KlCfD}) between $K_{l,C_l}$ and
$K_{l,D}$ to obtain the decomposition of $Z_j$ in terms of $K_{l,D}$. The
result is:
\begin{equation}
 Z_j=\sum_{l,D} b^{(l,D)}_j K_{l,D}
\label{expZj2}
\end{equation}
where the coefficients $b^{(l,D)}_j$ are given by
\begin{equation}
 b^{(l,D)}_j=\frac{b^{l}_j}{l!} +
 \sum_{(n_1>1) | l} \frac{b^{\left(\frac{l}{n_1},n_1>1\right)}_j}{l!} \;
 c\left(D,\left(\frac{l}{n_1},n_1\right)\right) \; .
\label{defblDj}
\end{equation}
Indeed, $K_{l}=\sum_{D} K_{l,D}$, and since $K_{(n_P,n_1)}$ corresponds to the
level $l=n_P n_1$, we have $K_{(n_P,n_1)}=\sum_{D \in S_{n_Pn_1}}
\frac{c(D,(n_P,n_1))}{l!} K_{n_P n_1,D}$. (Recall that $(n_P,n_1)$ is the
class of permutations composed of $n_P$ cycles of the same length $n_1$.) As
explained in Sec.~\ref{sec:coef_bl}, the $b^{(l,D)}_j$ are not simply equal to
$\frac{b^{l}_j}{l!}$ because of the $n_1>1$ terms. Using
Eq.~(\ref{relcDC}) we find that they nevertheless obey the following
relation
\begin{equation}
\sum_{D \in S_l} \left[{\rm dim}(D)\right]^2 b^{(l,D)}_j=b^{(l)}_j \; .
\label{relblDj}
\end{equation}
But from Eq.~(\ref{defblDj}) the $b^{(l,D)}_j$ with $l<2j$ are trivial, i.e.,
equal to $\frac{b^{l}_j}{l!}$ independently of $D$. This could have been shown
directly by considering the decomposition (\ref{expKltor}) of $K_l$. Finally,
since $b^{(1,n_1>1)}_1=0$ from Eq.~(\ref{defbnPn1>10}), only $b^{(l,D)}_0$ is
non-trivial for $l=2$ or $l=3$.

The decomposition of $Z$ over $K_{l,D}$ is obviously given by
\begin{equation}
Z=\sum_{l,D} b^{(l,D)} K_{l,D}
\label{expZKlD}
\end{equation}
where
\begin{equation}
b^{(l,D)}=\sum_{j=1}^l b^{(l,D)}_j \; .
\label{defblD}
\end{equation}
The $b^{(l,D)}$ then satisfy
\begin{equation}
\sum_{D \in S_l} \left[{\rm dim}(D)\right]^2 b^{(l,D)}=b^{(l)} \; .
\label{relblD2}
\end{equation}

\subsection{Relations among the $K_{l,D}$}

Just like the $K_{l,C_l}$, the $K_{l,D}$ are not independent, and for the same
reasons. Indeed, the number of $K_{l,D}$ which are independent among
themselves, and independent of $K_{l',D'}$ at higher levels $l'>l$, equals the
number of independent $K_{l,C_l}$. This number in turn equals the number of
integers dividing $l$, since the independent $K_{l,C_l}$ are $K_{l,{\rm Id}}$
and the $K_{(n_P,n_1)}$ with $l=n_P n_1$.

Therefore one can write relations between the $K_{l,D}$, by selecting an
independent number of $K_{l,D}$ and expressing the others in terms of those
selected. This produces relations of the form
\begin{equation}
K_{l,D_l}=\sum_{D_{l'}} e(D_l,D_{l'}) K_{l',D_{l'}}
\label{defe}
\end{equation}
where the $K_{l,D_l}$ are now those not selected, and the sum of
$K_{l',D_{l'}}$ is over the $D_{l'}$ selected with $l'\geq l$. The expressions
of the coefficients $e(D_l,D_{l'})$ depend of the choice of $K_{l,D}$ made.
Note in particular that to obtain the $e(D_l,D_{l'})$, the $K_{(n_P,n_1)'}$
must be expressed in terms of the $K_{(n_P,n_1)}$.
By combining Eqs.~(\ref{KnPn1'}) and (\ref{expZjn1}) we obtain
\begin{equation}
 K_{(n_P,n_1)'}=\sum_{n_P'=n_P+1}^{\left \lfloor \frac{L}{n_1} \right \rfloor}
 \sum_{j=n_P+1}^{n_P'}
 {j \choose n_P} {n_P' \choose j}
 (2^{j-n_P}-1)(-2)^{n_P'-j}
 K_{(n_P',n_1)} \;,
\end{equation}
and performing the sum over $j$ this becomes
\begin{equation}
 K_{(n_P,n_1)'}=\sum_{n_P'=n_P+1}^{\left \lfloor \frac{L}{n_1} \right \rfloor}
 {n_P' \choose n_P}
 (-1)^{n_P'-n_P+1}
 K_{(n_P',n_1)} \;.
\label{relKnPn1}
\end{equation}
Let us give an example of this relation: for $L=4$ we find that
$K_{3,(1,2)'}=2K_{4,(2,2)}$.

The coefficients $e(D_l,D_{l'})$ have the following properties:
\begin{eqnarray}
 \sum_{D_{l'} \in S_{l'=l}} \left[ {\rm dim}(D_{l'}) \right]^2 e(D_l,D_{l'})
 &=& \left[{\rm dim}(D_{l})\right]^2 \\
\sum_{D_{l'} \in S_{l'>l}} \left[ {\rm dim}(D_{l'}) \right]^2 e(D_l,D_{l'})
 &=& 0
\label{prope}
\end{eqnarray}
which can be proved using the fact that the $e(D_l,D_{l'})$ are independent of
$L$ and that the number of eigenvalues, including degeneracies, corresponding
to $K_{l,D_l}$ is $[{\rm dim}(D_{l})]^2 \, n_{\rm tor}(L,l)$. These relations
between $K_{l,D_l}$ have strong physical implications:  additional
degeneracies inside a level and between different levels. We shall give in the
next subsection a method to determine these degeneracies, but note for now
that they depend of $L$ although the $e(D_l,D_{l'})$ are independent of $L$.

We can now repeat the decompositions of the preceeding subsection, but
expanding only over the selected independent $K_{l,D}$. To that end,
we define the coefficients $\tilde{c}(D,C_l)$ by
\begin{equation}
 K_{l,C_l}= \sum_{{\rm indep.}\ D \in S_{l'\geq l}}
 \frac{\tilde{c}(D,C_l)}{(l')!} K_{l',D} \;.
\label{relKlCDind}
\end{equation}
Note that contrary to Eq.~(\ref{KlCfD}), the sum carries over all independent
(selected) $D$ at levels $l'\geq l$. Because of Eq.~(\ref{prope}), the
$\tilde{c}(D,C_l)$ have the following properties: if $C_l \neq {\rm Id}$ 
\begin{equation}
\sum_{{\rm indep.}\ D \in S_{l'}} [{\rm dim}(D)]^2 \tilde{c}(D,C_l) = 0 \;,
\label{propctilde}
\end{equation}
whereas if $C_l={\rm Id}$
\begin{eqnarray}
\sum_{{\rm indep.}\ D \in S_{l'=l}} [{\rm dim}(D)]^2 \tilde{c}(D,Id) &=&
 |C_l| \;, \nonumber \\
\sum_{{\rm indep.}\ D \in S_{l'>l}} [{\rm dim}(D)]^2 \tilde{c}(D,Id) &=& 0 \;.
\label{propctilde2}
\end{eqnarray}
Inserting the decomposition (\ref{relKlCDind}) of $K_{l,C_l}$ into
Eq.~(\ref{expZ}), we obtain the decomposition of $Z$ over independent
$K_{l,D}$:
\begin{equation}
Z=\sum_{l,D} \frac{\tilde{b}^{(l,D)}}{l!} K_{l,D}
\label{expZind}
\end{equation}
where the $\tilde{b}$ can be obtained using the $\tilde{c}$.

We do not have any general closed-form expression%
\footnote{The best one could hope for would be an explicit formula relating
 $\tilde{b}$ to the characters of the symmetric group.}
for $\tilde{b}$, but in the next subsection we show how they can be determined
in practice by a straightforward, though somewhat lengthy, procedure. More
precisely, we determine all the $\tilde{b}^{(l,D)}$ up to $l=4$, with a given
convention for the choice of independent $K_{l,D}$. As the $b^{(l,D)}$, the
$\tilde{b}^{(l,D)}$ verify
\begin{equation}
\sum_{{\rm indep.}\ D \in S_l} [{\rm dim}(D)]^2 \tilde{b}^{(l,D)} = b^{(l)}
\end{equation}
except that now the sum is over independent $D$. This is a consequence of the
properties of the $\tilde{c}$.

\subsection{Method to obtain the amplitudes of the eigenvalues}
\label{sec:amplitudes}

Because of the additional degeneracies between the $K_{l,D}$, we have not been
able to find a general formula giving the total degeneracies of the
eigenvalues. But, using Eq.~(\ref{expZ}) and the fact that the $c(D,C_l)$
defined by Eq.~(\ref{KlCfD}) are integers, we deduce that the amplitudes of the
eigenvalues are integer combinations of the $\frac{b^{(l)}}{l!}$ and the
$\frac{b^{(n_P,n_1>1)}}{(n_Pn_1)!}$. Determining precisely with which integers
is not an easy task, and we give here a method which is operational for all
values of $L$, though in practice it will probably become quite cumbersome for
large $L$.

One must begin at the highest possible level, $l=L$. Since not all
permutations are admissible, one can write relations between the $K_{L,D}$ and
deduce which eigenvalues are shared by several different $K_{L,D}$. One then
proceeds to the next lower-lying level, $l=L-1$. Since not all permutations
are admissible, and as some permutations are not independent of those at level
$l+1$, one can write relations between the $K_{l,D}$ and the $K_{l+1,D}$.
These relations permit to deduce which eigenvalues appearing at level $l$ are
new and what are their degeneracies. This method is then iterated until one
attains level $l=3$. Considering $l\leq 2$ is not necessary: all the
eigenvalues at these levels are new as there are no relations between the
corresponding $K_{l,D}$. Finally, using Eq.~(\ref{expZ}) where all $K_{l,D}$
have been expressed in terms of an independent number of $K_{l,D}$, we deduce
the amplitudes of the eigenvalues.

\Yboxdim4pt

Let us consider in detail the case $L=4$. At level $4$, the possible $K_{4,D}$
are $K_{\yng(4)}$, $K_{\yng(1,1,1,1)}$, $K_{\yng(3,1)}$, $K_{\yng(2,1,1)}$ and
$K_{\yng(2,2)}$, while the admissible $K_{l,C_l}$ are $K_{4,{\rm Id}}$,
$K_{4,(2,2)}$ and $K_{4,(1,4)}$. Using the table of characters of $S_4$, we
can write:
\begin{eqnarray}
K_{\yng(4)} &=& K_{4,{\rm Id}}+K_{4,(2,2)}+K_{4,(1,4)} \label{chtabK4} \\
K_{\yng(1,1,1,1)} &=& K_{4,{\rm Id}}+K_{4,(2,2)}-K_{4,(1,4)} \label{chtabK1111} \\
K_{\yng(3,1)} &=& 9K_{4,{\rm Id}}-3K_{4,(2,2)}-3K_{4,(1,4)} \\
K_{\yng(2,1,1)} &=& 9K_{4,{\rm Id}}-3K_{4,(2,2)}+3K_{4,(1,4)} \\
K_{\yng(2,2)} &=& 4K_{4,{\rm Id}}+4K_{4,(2,2)}
\end{eqnarray}
We choose $K_{\yng(4)}$, $K_{\yng(3,1)}$ and $K_{\yng(2,2)}$ as independent
$K_{4,D}$, and we express the $K_{4,C_l}$ in terms of those $K_{4,D}$:
\begin{eqnarray}
K_{4,{\rm Id}}&=&\frac{K_{\yng(4)}}{4}+\frac{K_{\yng(3,1)}}{12} \\
K_{4,(2,2)}&=&-\frac{K_{\yng(4)}}{4}-\frac{K_{\yng(3,1)}}{12}+\frac{K_{\yng(2,2)}}{4} \\
K_{4,(1,4)}&=&K_{\yng(4)}-\frac{K_{\yng(2,2)}}{4} 
\end{eqnarray}
Next, using these expressions, we obtain $K_{\yng(1,1,1,1)}$ and
$K_{\yng(2,1,1)}$ in terms of the independent $K_{4,D}$ chosen:
\begin{eqnarray}
K_{\yng(1,1,1,1)}&=&-K_{\yng(4)}+\frac{K_{\yng(2,2)}}{2} \label{relK1111} \\
K_{\yng(2,1,1)}&=&6K_{\yng(4)}+K_{\yng(3,1)}-\frac{3}{2}K_{\yng(2,2)}
\label{relK211}
\end{eqnarray}
With these two relations we can determine the eigenvalue degeneracies between
the chosen $K_{4,D}$. Recall first that according to Eq.~(\ref{defastrace})
the number of eigenvalues contributing to $K_{l,D}$ is $n_{\rm tor}(L,l) \,
{\rm dim}(D)$ and that each eigenvalue has multiplicity ${\rm dim}(D)$.
Further, ${\rm dim}(\yng(4))=1$, ${\rm dim}(\yng(3,1))=3$ and ${\rm
dim}(\yng(2,2))=2$. Consider now Eq.~(\ref{relK1111}), recalling that the
$K_{l,D}$ have been defined in Eq.~(\ref{defastrace}) as a trace. We deduce
that the corresponding eigenvalues must satisfy
\beq \left( \lambda_{\yng(1,1,1,1)} \right)^N = -
 \left(\lambda_{\yng(4)}\right)^N + \sum_{i=1}^2
 \left(\lambda_{\yng(2,2),i}\right)^N \eeq
for any positive integer $N$. This implies that $\lambda_{\yng(2,2),1} =
\lambda_{\yng(4)}$ and that $\lambda_{\yng(2,2),2} = \lambda_{\yng(1,1,1,1)}$.
Using this, Eq.~(\ref{relK211}) then yields
\beq
 \sum_{i=1}^3 \left(\lambda_{\yng(2,1,1),i}\right)^N =
 \left(\lambda_{\yng(4)}\right)^N - \left(\lambda_{\yng(1,1,1,1)}\right)^N +
 \sum_{i=1}^3 \left(\lambda_{\yng(3,1),i}\right)^N
\eeq
for any $N$. This is possible provided that either
$\lambda_{\yng(1,1,1,1)}=\lambda_{\yng(4)}$ or
$\lambda_{\yng(1,1,1,1)}=\lambda_{\yng(3,1),1}$. But the first possibility can
be excluded since, by Eqs.~(\ref{chtabK4})--(\ref{chtabK1111}), it would imply
$K_{4,(1,4)}=0$ which is inconsistent with our hypothesis that we work on a
generic lattice where all admissible $K$ are non-zero. We conclude that
$\lambda_{\yng(1,1,1,1)}=\lambda_{\yng(3,1),1}$ and hence that
$\lambda_{\yng(4)}=\lambda_{\yng(2,1,1),1}$ and $\lambda_{\yng(3,1),i} =
\lambda_{\yng(2,1,1),i}$ for $i=2,3$. There are therefore only $4$ different
eigenvalues at level $4$ instead of $10$.

Consider now the level $l=3$. From the character table of $S_3$ we obtain:
\begin{eqnarray}
K_{\yng(3)} &=& K_{3,{\rm Id}}+K_{3,(1,2)'}+K_{3,(1,3)} \\
K_{\yng(1,1,1)} &=& K_{3,{\rm Id}}-K_{3,(1,2)'}+K_{3,(1,3)} \\
K_{\yng(2,1)} &=& 4K_{3,{\rm Id}}-2K_{3,(1,3)}
\end{eqnarray}
$K_{3,(1,2)'}$ must be expressed terms of the independent $K_{4,D_l}$
chosen:
\begin{equation}
K_{3,(1,2)'}=2K_{4,(2,2)}=-\frac{K_{\yng(4)}}{2}-\frac{K_{\yng(3,1)}}{6}+\frac{K_{\yng(2,2)}}{2}
\end{equation}
We choose $K_{\yng(3)}$ and $K_{\yng(2,1)}$ as independent $K_{3,D}$,
and we express the $K_{3,C_l}$ in terms of the independent $K_{3,D}$
and $K_{4,D}$ chosen:
\begin{eqnarray}
 K_{3,{\rm Id}}&=&\frac{K_{\yng(3)}}{3}+\frac{K_{\yng(2,1)}}{6}+\frac{K_{\yng(4)}}{6}+\frac{K_{\yng(3,1)}}{18}-\frac{K_{\yng(2,2)}}{6} \\
 K_{3,(1,3)}&=&\frac{2}{3}K_{\yng(3)}-\frac{K_{\yng(2,1)}}{6}+\frac{K_{\yng(4)}}{3}+\frac{K_{\yng(3,1)}}{9}-\frac{K_{\yng(2,2)}}{3}
\end{eqnarray}
We obtain then the expression of $K_{\yng(1,1,1)}$:
\begin{equation}
K_{\yng(1,1,1)}=K_{\yng(3)}+K_{\yng(4)}+\frac{K_{\yng(3,1)}}{3}-K_{\yng(2,2)}
\label{Kidlevel3}
\end{equation}
Using again Eq.~(\ref{defastrace}) and the eigenvalue identities obtained
at level $l=4$, this becomes
\beq
 \sum_{i=1}^8 \left(\lambda_{\yng(1,1,1),i}\right)^N =
 \sum_{i=1}^8 \left(\lambda_{\yng(3),i}\right)^N -
 \left( \lambda_{\yng(4)} \right)^N -
 \left( \lambda_{\yng(3,1),1} \right)^N +
 \left( \lambda_{\yng(3,1),2} \right)^N +
 \left( \lambda_{\yng(3,1),3} \right)^N
\eeq
from which we deduce that $\lambda_{\yng(3),1} = \lambda_{\yng(4)}$, and that
$\lambda_{\yng(3),2} = \lambda_{\yng(3,1),1}$. (Note that we cannot have, for
example, $\lambda_{\yng(4)} = \lambda_{\yng(3,1),2}$ since these eigenvalues
were shown to be independent in the preceeding analysis at level $l=4$.) We
can then further deduce that $\lambda_{\yng(1,1,1),1} =
\lambda_{\yng(3,1),2}$, that $\lambda_{\yng(1,1,1),2} =
\lambda_{\yng(3,1),3}$, and that $\lambda_{\yng(1,1,1),i} =
\lambda_{\yng(3),i}$ for $i=3,4,\ldots,8$. So among the 8 eigenvalues
participating in $K_{\yng(3)}$ and the 8 participating in $K_{\yng(1,1,1)}$
only a total of 6 are new. On the other hand, all $16$ eigenvalues
participating in $K_{\yng(2,1)}$ are new, since $K_{\yng(2,1)}$ did not appear
in an identity such as Eq.~(\ref{Kidlevel3}).

The eigenvalues for $l\leq 2$ are all new, as there are no relations between
the $K_{l,D}$. Therefore, there are $28$ new eigenvalues associated to
$K_{\yng(2)}$, $28$ to $K_{\yng(1,1)}$, $35$ to $K_1$ and $14$ to $K_0$.

To obtain the amplitudes associated to the eigenvalues, we use
Eq.~(\ref{expZ}):
\begin{eqnarray}
Z&=&K_0 + b^{(1)} K_1 + b^{(2)} K_{2,{\rm Id}} + b^{(1,n_1>1)} K_{2,(1,2)} +
b^{(3)} K_{3,{\rm Id}} + b^{(1,n_1>1)} K_{3,(1,3)} \nonumber \\
&+& b^{(4)} K_{4,{\rm Id}} + b^{(1,n_1>1)} K_{4,(1,4)} + b^{(2,n_1>1)}
K_{4,(2,2)}
\end{eqnarray}
and insert the expressions of the $K_{l,C_l}$ in terms of the independent
$K_{l,D}$ chosen above:
\begin{eqnarray}
Z&=&K_0 + b^{(1)} K_1 + \tilde{b}^{(\yng(2))}K_{\yng(2)}+ \tilde{b}^{(\yng(1,1))}K_{\yng(1,1)} + \tilde{b}^{(\yng(3))}K_{\yng(3)}+ \tilde{b}^{(\yng(2,1))}K_{\yng(2,1)} \nonumber \\
&+&\tilde{b}^{(\yng(4))}K_{\yng(4)} + \tilde{b}^{(\yng(3,1))}K_{\yng(3,1)} + \tilde{b}^{(\yng(2,2))}K_{\yng(2,2)} \label{expZforL4}
\end{eqnarray}
where the amplitudes associated to the independent $K_{l,D}$ are:
\begin{eqnarray}
\tilde{b}^{(\yng(2))}&=&\frac{b^{(2)}}{2}+\frac{b^{(1,n_1>1)}}{2} \\
\tilde{b}^{(\yng(1,1))}&=&\frac{b^{(2)}}{2}-\frac{b^{(1,n_1>1)}}{2} \\
\tilde{b}^{(\yng(3))}&=&\frac{b^{(3)}}{3}+\frac{2 b^{(1,n_1>1)}}{3} \\
\tilde{b}^{(\yng(2,1))}&=&\frac{b^{(3)}}{6}-\frac{b^{(1,n_1>1)}}{6} \\
\tilde{b}^{(\yng(4))}&=&\frac{b^{(3)}}{6}+\frac{4b^{(1,n_1>1)}}{3}+\frac{b^{(4)}}{4}-\frac{b^{(2,n_1>1)}}{4}\\
\tilde{b}^{(\yng(3,1))}&=&\frac{b^{(3)}}{18}+\frac{b^{(1,n_1>1)}}{9}+\frac{b^{(4)}}{12}-\frac{b^{(2,n_1>1)}}{12} \\
\tilde{b}^{(\yng(2,2))}&=&-\frac{b^{(3)}}{6}-\frac{7b^{(1,n_1>1)}}{12}+\frac{b^{(2,n_1>1)}}{4}
\end{eqnarray}
These can now be calculated explicitly from Eqs.~(\ref{defbltor})
and (\ref{defbnPn1>1}):
\begin{eqnarray}
\tilde{b}^{(\yng(2))}&=&\frac{Q^2}{2}-\frac{3Q}{2} \\
\tilde{b}^{(\yng(1,1))}&=&\frac{Q^2}{2}-\frac{3Q}{2}+1 \\
\tilde{b}^{(\yng(3))}&=&\frac{Q^3}{3}-2Q^2+\frac{8Q}{3}-1 \\
\tilde{b}^{(\yng(2,1))}&=&\frac{Q^3}{6}-Q^2+\frac{4Q}{3} \\
\tilde{b}^{(\yng(4))}&=&\frac{Q^4}{4}-\frac{11Q^3}{6}+\frac{15Q^2}{4}-\frac{5Q}{3}-2 \\
\tilde{b}^{(\yng(3,1))}&=&\frac{Q^4}{12}-\frac{11Q^3}{18}+\frac{15Q^2}{12}-\frac{5Q}{9}-\frac{1}{3} \\
\tilde{b}^{(\yng(2,2))}&=&-\frac{Q^3}{6}+\frac{5Q^2}{4}-\frac{25Q}{12}+\frac{3}{2}
\end{eqnarray}
Note that the four first amplitudes in this list have been obtained by Chang
and Schrock \cite{shrock} using a different method.

We can finally give the amplitudes of the eigenvalues themselves. The $14$
eigenvalues at level $0$ have amplitude $1$. The $35$ eigenvalues at level $1$
have amplitude $\tilde{b}^{(1)}$. At level $2$, the $24$ eigenvalues
contributing to $K_{\yng(2)}$ have amplitude $\tilde{b}^{\yng(2)}$, and the
$24$ eigenvalues contributing to $K_{\yng(1,1)}$ have amplitude
$\tilde{b}^{\yng(1,1)}$. At level $3$, the $6$ new eigenvalues contributing to
$K_{\yng(3)}$ have amplitude $\tilde{b}^{\yng(3)}$, and the $16$ eigenvalues
contributing to $K_{\yng(2,1)}$ have amplitude $2\tilde{b}^{\yng(2,1)}$. At
level $4$, $\lambda_{\yng(4)}$ has amplitude
$\tilde{b}^{\yng(4)}+2\tilde{b}^{\yng(2,2)}+\tilde{b}^{\yng(3)}$,
$\lambda_{\yng(1,1,1,1)}$ has amplitude
$3\tilde{b}^{\yng(3,1)}+2\tilde{b}^{\yng(2,2)}+\tilde{b}^{\yng(3)}$, and
$\lambda_{\yng(3,1),1}$ and $\lambda_{\yng(3,1),2}$ both have the same
amplitude $3\tilde{b}^{\yng(3,1)}$.

Note that when we know the amplitudes of the $K_{l,D}$ in the expansion of $Z$
and the relations between $K_{l,D}$ for a given width $L$, we know it for all
widths smaller than $L$, as they do not change. The only difference is that
for smaller widths some of the $K_{l,D}$ vanish, so the equations must be
truncated. For example, the result (\ref{Kidlevel3}) obtained here for $L=4$,
implies by truncation of the $K_{4,D}$ terms that for $L=3$:
\begin{equation}
K_{\yng(1,1,1)}=K_{\yng(3)}
\end{equation}
Likewise, the expansion (\ref{expZforL4}) of $Z$ obtained here for $L=4$,
implies by truncation that for $L=3$:
\begin{equation}
Z=K_0 + b^{(1)} K_1 + \tilde{b}^{(\yng(2))}K_{\yng(2)}+ \tilde{b}^{(\yng(1,1))}K_{\yng(1,1)}+ \tilde{b}^{(\yng(3))}K_{\yng(3)}+ \tilde{b}^{(\yng(2,1))}K_{\yng(2,1)}
\end{equation}
From these equations, it is then simple to obtain the amplitudes of
eigenvalues for $L=3$.

\subsection{Particular non-generic lattices}

The degeneracies we have obtained apply to the case of a generic lattice. In
the case of a specific lattice, i.e., one having extra non-generic symmetries,
there might be additional degeneracies. An example is the case of a square or
a honeycomb lattice, because of the invariance of the lattice under reflection
by its symmetry axis.

Specifically, the transfer matrix ${\rm T}_L$ of the square lattice commutes
with the dihedral group $D_L$, since the lattice enjoys both translational and
reflectional symmetries in the space perpendicular to the transfer direction.
Likewise, for the honeycomb lattice, the symmetry group is $D_{\frac{L}{2}}$.
Those groups act on the right time slice, not on the left one (i.e., the black
points), and thus commute with the symmetry group $S_l$ at level $l$ of the
bridges. There are therefore additional degeneracies inside a given $K_{l,D}$.

In the case of the dihedral group, since all its irreps are of dimension $1$
or $2$, there are additional degeneracies between pairs of eigenvalues, as
observed in Ref.~\cite{shrock}. The method to determine precisely these
degeneracies is to decompose the space at level $l$ of symmetry $D$ into
irreps of the dihedral group, and to count the number of irreps of dimension
$2$.

Furthermore, in the case of a square or a hexagonal lattice, there are yet
additional degeneracies (which do not exist for a generic lattice with a
dihedral symmetry). Indeed, for these two lattices, the $Z_{L,n_1}$ with
$n_1>1$ are found to be zero: at level $l=L$ no permutation between bridges is
allowed. At this level there is thus only one eigenvalue of total degeneracy
$\sum_{D \in S_l} \left[ {\rm dim}(D)\right]^2 \tilde{b}^{(L,D)}=b^{(L)}$.
Finally, in the case of a square lattice, there appears to be additional
``accidental'' degeneracies: for example, an eigenvalue at level $1$ coincides
with an eigenvalue at level $2$, as observed in Ref.~\cite{shrock}.

\section{Conclusion}

To summarise, we have generalised the combinatorial approach developed in
Ref.~\cite{cyclic} for cyclic boundary conditions to the case of toroidal
boundary conditions. In particular, we have obtained the decomposition of the
partition function for the Potts model on finite tori in terms of the
generalised characters $K_{l,D}$. This decomposition is considerably more
difficult to interpret than in the cyclic case, as some eigenvalues coincide
between different levels $l$ for all values of $Q$. We have nevertheless
succeeded in giving an operational method of determining the amplitudes of the
eigenvalues as well as their {\em generic} degeneracies.

The eigenvalue amplitudes are instrumental in determining the physics of the
Potts model, in particular in the antiferromagnetic regime \cite{af}.
Generically, this regime belongs to a so-called Berker-Kadanoff (BK) phase in
which the temperature variable is irrelevant in the renormalisation group
sense, and whose properties can be obtained by analytic continuation of the
well-known ferromagnetic phase transition \cite{hs}. Due to the
Beraha-Kahane-Weiss (BKW) theorem \cite{bkw}, partition function zeros
accumulate at the values of $Q$ where either the amplitude of the dominant
eigenvalue vanishes, or where the two dominant eigenvalues become equimodular.
When this happens, the BK phase disappears, and the system undergoes a phase
transition with control parameter $Q$. Determining analytically the eigenvalue
amplitudes is thus directly relevant for the first of the hypotheses in the
BKW theorem.

For the cyclic geometry, the amplitudes are very simple, and the values of $Q$
satisfying the hypothesis of the BKW theorem are simply the so-called Beraha
numbers, $Q=B_n=(2 \cos(\pi/n))^2$ with $n=2,3,\ldots$, independently of the
width $L$. For the toroidal case, we have no general formula for the
amplitudes, valid for any $L$. It is however clear from the amplitudes given
for $L \le 4$ in Sec.~\ref{sec:amplitudes} that many of them vanish at $Q=2$,
and yet other differ just by a sign by virtue of Eq.~(\ref{defbnPn1>1}).
Indeed, it is consistent with simple physical arguments, that a phase
transition in the antiferromagnetic regime must take place at $Q=2$. However,
it remains to elucidate whether the BK phase exists for all other $Q \in
(0,4)$, and whether the Beraha numbers play any special role in the toroidal
case.

\vspace{0.1cm}

\noindent
{\bf Acknowledgments.}

The authors are grateful to P.~Zinn-Justin and J.-B.~Zuber for some useful
discussions. JLJ further thanks the members of the SPhT, where part of this
work was done, for their kind hospitality.

\end{document}